\documentclass[letterpaper, 10 pt, conference]{ieeeconf}  

\IEEEoverridecommandlockouts                              
\usepackage{epsfig} 
\usepackage{amsmath,amssymb,amsfonts}
\usepackage{color}
\usepackage{balance}
\usepackage{booktabs,tabularx}
\usepackage{multirow}
\usepackage{mathtools}
\usepackage{pdfrender}
\usepackage{trfsigns}

\usepackage{centernot}
\usepackage{color}
\usepackage[acronym]{glossaries}
\usepackage[ruled]{algorithm2e}
\usepackage{graphicx,subcaption}
\usepackage{tabu}
\usepackage{cite}
\usepackage[official]{eurosym}

\graphicspath{{figures/}}

\newcommand{\R}{\mathbb{R}}

\newcommand{\N}{\mathbb{N}}

\newcommand{\mc}[1]{\mathcal{#1}}

\newcommand{\prob}{\mathbb{P}}

\newcommand{\bs}{\boldsymbol}
\newcommand{\bsone}{\boldsymbol{1}}

\newcommand{\col}{\mathrm{col}}

\newtheorem{definition}{Definition}
\newtheorem{proposition}{Proposition}
\newtheorem{lemma}{Lemma}

\newtheorem{remark}{Remark}

\newtheorem{standing}{Standing Assumption}

\usepackage{xcolor,calc}

\makeglossaries
\makeglossaries
\newacronym{NEP}{NEP}{Nash equilibrium problem}
\newacronym{GNEP}{GNEP}{generalized Nash equilibrium problem}
\newacronym{iid}{i.i.d.}{independent and identically distributed}
\newacronym{w.r.t.}{w.r.t.}{with respect to}
\newacronym{od}{OD}{origin-destination}
\newacronym{ncp}{NCP}{nonlinear complementarity problem}
\newacronym{LTI}{LTI}{linear time-invariant}
\newacronym{PEV}{PEV}{plug-in electric vehicle}
\newacronym{SoC}{SoC}{State of Charge}

\newglossaryentry{QVI}
{
	name={VI},
	description={quasi variational inequality},
	first={\glsentrydesc{QVI} (\glsentrytext{QVI})},
	plural={VIs},
	descriptionplural={quasi variational inequalities},
	firstplural={\glsentrydescplural{QVI} (QVIs)}
}

\newglossaryentry{VI}
{
	name={VI},
	description={variational inequality},
	first={\glsentrydesc{VI} (\glsentrytext{VI})},
	plural={VIs},
	descriptionplural={variational inequalities},
	firstplural={\glsentrydescplural{VI} (VIs)}
}

\newglossaryentry{v-GNE}
{
	name={v-GNE},
	description={variational generalized Nash equilibrium},
	first={\glsentrydesc{v-GNE} (\glsentrytext{v-GNE})},
	plural={v-GNE},
	descriptionplural={variational generalized Nash equilibria},
	firstplural={\glsentrydescplural{v-GNE} (\glsentryplural{v-GNE})}
}

\newglossaryentry{GNE}
{
	name={GNE},
	description={generalized Nash equilibrium},
	first={\glsentrydesc{GNE} (\glsentrytext{GNE})},
	plural={GNE},
	descriptionplural={generalized Nash equilibria},
	firstplural={\glsentrydescplural{GNE} (\glsentryplural{GNE})}
}

\title{\LARGE \bf
Pursuing robust decisions in uncertain traffic equilibrium problems
}

\author{Filippo Fabiani
\thanks{The author is with the Department of Engineering Science, University of Oxford, OX1 3PJ, United Kingdom {\tt \footnotesize (filippo.fabiani@eng.ox.ac.uk)}. This work was partially supported through the Government’s modern industrial strategy by Innovate UK, part of UK Research and Innovation, under Project LEO (Ref. 104781).}%
}

\begin{document}

\maketitle
\thispagestyle{empty}
\pagestyle{empty}

\begin{abstract}

We evaluate the robustness of agents' traffic equilibria in randomized routing games characterized by an uncertain network demand with a possibly unknown probability distribution. Specifically, we extend the so-called \emph{hose model} by considering a traffic equilibrium model where the uncertain network demand configuration belongs to a polyhedral set, whose shape is itself a-priori unknown.
By exploiting available data, we apply the scenario approach theory to establish distribution-free feasibility guarantees for agents' traffic equilibria of the uncertain routing game without the need to know an explicit characterization of such set.
A numerical example on a traffic network testbed corroborates the proposed theoretical results.

\end{abstract}

\section{Introduction}
Tracing back to the 70s, traffic routing problems essentially consist of carrying traffic from origins to destinations by making use of the (typically limited) network resources \cite{smith1979existence,dafermos1980traffic}. Additionally, the routing pattern shall also satisfy some quality of service constraints, as well as intrinsic physical limitations of the communication infrastructures.
 
In this framework, congestion is a recurring phenomenon observed in several domains, spanning from large urban areas \cite{florian1995network,patriksson1994traffic} to telecommunication networks \cite{gojmerac2003adaptive,chiang2007layering}. 
The circulation flow associated to congested networks is popularly described in a static fashion by means of traffic equilibrium models or nonatomic routing games, where the number of users is assumed to be large, each one controlling an infinitesimal amount of flow in the network (with a slight abuse of terminology, we will make use of the two models interchangeably). 
Here, an equilibrium typically emerges as a steady-state from the adaptive behaviour of selfish agents, which strive to minimize their own travel time or transportation cost, while sharing the limited network resources. A widely diffused stationarity notion, i.e., the Wardrop equilibrium \cite{wardrop1952road}, establishes that origin-destination paths with non-zero flow have the least cost among all the possible alternatives, thus enabling the equivalence between path flow equilibria and solutions to the \gls{VI} associated with the routing game \cite[Th.~3.14]{patriksson1994traffic}.

Congestion models can then be adopted to forecast flows, in order to evaluate either the overall network performance under different scenarios of demand, or alternative traffic management policies. In real-life, however, path flows and network travel demands are often variable over time in a non-regular and unpredictable manner. Such an uncertainty can not only be caused by a particular hour of the day, but also by sudden accidents or maintenance works. These reasons essentially motivate the need for uncertain traffic and network models, as thoroughly described in, e.g.,  \cite{fingerhut1997designing,ben2005routing,gwinner2006random,ouorou2007model,lemarechal2010robust,frangioni2011static,ouorou2013tractable,daniele2015random,cominetti2015equilibrium,jadamba2018efficiency,cherukuri2019sample}.

Following such a literature body, this paper considers a routing problem where a certain source of uncertainty affects the network traffic demands, and studies the uncertain \gls{VI} that equivalently expresses the Wardrop equilibrium conditions. Specifically, to capture the traffic variations that typically generate an a-priori unpredictable network demand, we adopt the approach proposed in \cite{ben2005routing}, thus constraining the admissible traffic configurations to fall into a \textit{traffic demand polyhedron}. Allowing one to bypass traditional probabilistic assumptions on the network demand as postulated in, e.g., \cite{gwinner2006random,daniele2015random,cominetti2015equilibrium,jadamba2018efficiency}, such a traffic demand polyhedron generalizes the so-called \textit{hose model} \cite{fingerhut1997designing}, and has been largely adopted in both static and dynamic routing problems \cite{ouorou2007model,lemarechal2010robust,frangioni2011static,ouorou2013tractable}. 
However, since making a-priori traffic predictions is quite challenging, we note that it might be unlikely to have available an explicit collection of deterministic linear inequalities describing all of the relevant traffic demand configurations.

For this reason, we accommodate a certain degree of uncertainty  on the actual shape of the traffic demand polyhedron by parametrizing such a set with a random variable. 
In this way, we implicitly extend the approach in \cite{ben2005routing} by providing an additional degree of freedom in evaluating the network performance under different demand scenarios or traffic management policies. 
We then investigate a randomized approach to a-posteriori quantify the feasibility risk associated to any agents' traffic equilibrium of the randomized routing game against unseen realizations of the uncertain parameter. 
Specifically, we leverage the equivalence between routing games and \glspl{VI} \cite[Th.~3.14]{patriksson1994traffic} to rely on recent results bridging the realm of the \glspl{VI} with the scenario approach paradigm \cite{paccagnan2019scenario,fabiani2020probabilistic}, with the twofold benefit of enabling for a tractable reformulation of the uncertain routing game, and establishing quantifiable robustness properties for any of the agents' traffic equilibria in a distribution-free fashion. That is, the proposed robustness certificates characterizing randomized optimal routing patterns hold regardless of the probability distribution of the parameter that encodes the uncertainty on the shape of the traffic demand polyhedron.

The paper is organized as follows: in \S II, we introduce the notation and recall some concepts of graph theory and \glspl{VI}, while we describe in detail the uncertain traffic model adopted in \S III. In \S IV, we formalize the scenario-based traffic equilibrium problem, and in \S V we provide the robustness certificates for traffic equilibria. Finally, we corroborate our theoretical findings through a numerical example in \S VI.


\section{Notation and Preliminaries}
We start by introducing the notation and some key ingredients of graph theory and \glspl{VI} adopted throughout the paper.

\subsubsection{Notation} 
$\N$, $\R$, $\R_{> 0}$ and $\R_{\geq 0}$ denote the set of natural, real, positive real, nonnegative real numbers, respectively. $\N_0 \coloneqq \N \cup \{0\}$. 
$\bs{1}$ ($\bs{0}$) denotes vectors of appropriate dimensions with elements all equal to $1$ ($0$). Given a matrix $A \in \R^{m \times n}$, its $(i,j)$ entry is denoted by $a_{i,j}$, $A^\top$ denotes its transpose.
For a set $\mc{X} \subseteq \R^n$, $|\mc{X}|$ represents its cardinality. $\mathrm{int}(\mc{X})$
and $\mathrm{bdry}(\mc{X})$ denote its topological interior
and boundary, respectively; $\mathrm{aff}(\mc{X})$ denotes its affine hull, i.e., the smallest affine set containing $\mc{X}$. 
The mapping $T: \mc{X} \rightarrow \R^n$ is monotone if $(T(x) - T(y))^\top(x - y) \geq  0$ for all $x, y \in \mc{X}$. If $\mc{X}$ is nonempty and convex, the normal cone of $\mc{X}$ evaluated at  $x$ is the multi-valued mapping $\mc{N}_{\mc{X}} : \R^n \rightrightarrows \R^n$, defined as $\mc{N}_{\mc{X}}(x) \coloneqq \{	d \in \R^n \mid d^\top (y - x) \leq 0, \; \forall y \in \mc{X}	\}$ if $x \in \mc{X}$, $\mc{N}_{\mc{X}}(x) \coloneqq \emptyset$ otherwise.
The operator $\col(\cdot)$ stacks in column vectors or matrices of compatible dimensions, while $\mathrm{avg}(\cdot)$ returns the average among the elements in the argument.

\subsubsection{Graph theory (\hspace{-.03mm}\textup{\cite{mesbahi2010graph}})}
A \textit{directed graph} (or digraph) is a pair $\mc{G} \coloneqq (\mc{V}, \mc{E})$, where $\mc{V}$ denotes the finite set of \textit{nodes} (or vertices), while $\mc{E} \subseteq \mc{V} \times \mc{V}$ the set of \textit{edges} (or arcs), and $(i,j) \in \mc{E}$ if there exists an oriented edge from node $i$ to $j$. A \textit{directed path} is a sequence of distinct nodes such that any two subsequent nodes form a directed edge. A digraph is \textit{strongly connected} if, for every pair of vertices, there exists a directed path between them. A \textit{source} is a node with no incoming edge, while a \textit{sink} is a node with no outgoing edge.

\subsubsection{Variational inequality (\hspace{-.03mm}\textup{\cite{facchinei2007finite}})} 
Formally, a \gls{VI} is defined by means of a feasible set $\mc{X} \subseteq \R^n$, and a mapping $F : \mc{X} \to \R^n$. We denote by VI$(\mc{X}, F)$ the problem of finding some vector $x^\star \in \mc{X}$ such that $(y - x^\star)^\top F(x^\star) \geq 0, \, \text{ for all } y \in \mc{X}$. Such an $x^\star$ is therefore called a \textit{solution} to VI$(\mc{X}, F)$, and the associated set of solutions is denoted as $\mc{S} \subseteq \mc{X}$.

\section{Model description}
Inspired by \cite{ben2005routing,ouorou2007model,lemarechal2010robust,ouorou2013tractable,cominetti2015equilibrium,jadamba2018efficiency}, we first introduce the uncertain traffic model adopted, and then we describe the robust decision-making problem that follows.

\subsection{Routing games with uncertain traffic demand}
A traffic network can be formally described through a directed, strongly connected graph $\mc{G} \coloneqq (\mc{V}, \mc{E})$, in which we denote $\mc{L} \subseteq \mc{V} \times \mc{V}$ as the set indexing all \gls{od} pairs among the nodes of the network, $\ell \coloneqq |\mc{L}|$. Specifically, origin and destination nodes coincide with the sources and sinks in the network, respectively. In view of the strong connectivity of $\mc{G}$, each \gls{od} pair is connected by at least one path, and we therefore denote with $\mc{M}$ the set indexing all the paths in the network, with $m \coloneqq |\mc{M}|$.

We assume that a large number of agents aims at crossing the network in a noncooperative manner. Specifically, we suppose each agent being associated with an \gls{od} pair, and it is allowed to select any path $r \in \mc{M}$ connecting such a pair. This framework is modelled as a nonatomic routing game where each individual agent’s action has an infinitesimal impact on the aggregate traffic flow. As a consequence, the flow on the $i$-th edge, $f_i \geq 0$, $i \in \mc{E}$, is a continuous variable.

The arc-path incidence matrix $B \in \R^{e \times m}$, $e \coloneqq |\mc{E}|$, associated with $\mc{G}$ allows us to describe the edge structure of the paths. In fact, for all $(i,j) \in \mc{E} \times \mc{M}$, we have that
\begin{equation}\label{eq:B_def}
	b_{i,j} \coloneqq \left\{
	\begin{aligned}
		& 1 \quad \text{ if arc $i \in \mc{E}$ belongs to path $j \in \mc{M}$, }\\
		& 0 \quad \text{ otherwise. }
	\end{aligned}
	\right.
\end{equation}
Then, it follows that to each path $r \in \mc{M}$ corresponds a nonnegative flow $p_r \geq 0$, stacked together in $\bs{p} \coloneqq \col((p_r)_{r \in \mc{M}}) \in \R_{\geq 0}^m$, which is confined in some set $\mc{P}$. In particular, the constraint set $\mc{P}$ rules out the unrealistic case that each path can have either negative or infinite capacity, and it is hence typically formalized as
$
\mc{P} \coloneqq \{\bs{p} \in \R^m \mid \bs{p} \in [\bs{0}, \bs{p}_{\textrm{max}}] \},
$
for some given $\bs{p}_{\textrm{max}} > 0$. Note that the flow $f_i$ on the $i$-th edge is equal to the sum of the path flows on the paths that contain the $i$-th edge, so that we have $\bs{f} = B \bs{p}$. 

Associated with the set of \gls{od} pairs $\mc{L}$, and therefore with the route choice of all the agents, an aggregate traffic demand is given that shall satisfy the overall network demand. Specifically, in view of its a-priori unpredictability, such a network demand is here modelled through an (uncertain) \emph{traffic demand polyhedron} \cite{ben2005routing}.
Let us now introduce $H \in \R^{\ell \times m}$ as the \gls{od} pair-path incidence matrix whose generic entry $h_{i,j}$ is equal to $1$ if the path $j \in \mc{M}$ connects the pair $i \in \mc{L}$, $0$ otherwise. Then, the aggregate traffic demand $H \bs{p} \in \R^{\ell}_{\geq 0}$ shall belong to the following uncertain polyhedral set, which is parametrized by a random variable $\omega \in  \R^d$,
\begin{equation}\label{eq:unc_set}
	\mc{P}_\omega \coloneqq \{\bs{p} \in \R^{m} \mid A(\omega) H \, \bs{p} \leq b(\omega)\} \cap \mc{P}, \, \omega \in \Omega,
\end{equation}
where $A : \R^d \to \R^{s \times \ell}$ and $b : \R^d \to \R^s$. We assume the uncertain parameter $\omega$ defined over the probability space $(\Omega,\mc{D},\mathbb{P})$, where $\Omega \subseteq \R^d$ represents the set of values that $\omega$ can take, $\mc{D}$ is the associated $\sigma$-algebra and $\mathbb{P}$ is a (possibly unknown) probability measure over $\mc{D}$. The parameter $\omega$ reflects the fact that making a-priori traffic predictions is quite challenging, and hence it might be difficult to have available a collection of deterministic linear inequalities describing all of the relevant traffic demand configurations as postulated, e.g., in \cite{ben2005routing,lemarechal2010robust,ouorou2013tractable,ouorou2007model}. With the relation in \eqref{eq:unc_set}, indeed, we not only assume that the traffic demand is uncertain, thus belonging to a (possibly unbounded) polyhedral set, but also the shape of such a set is a-priori unknown. This latter source of uncertainty is hence encoded by the random parameter $\omega$.

%

We now introduce the unit cost of going through the edge $i \in \mc{E}$ as a nonnegative function $c_i : \R_{\geq 0}^m \to \R_{\geq 0}$ of the overall flows on the network, so that $c: \R_{\geq 0}^{m} \to \R_{\geq 0}^{e}$, defined as $c(B\bs{p}) \coloneqq \col((c_i( B\bs{p} ))_{i \in\mc{E}})$ denotes the arc cost vector of the network. Here, every function $c_i(\cdot)$ is usually associated with the travel time or transportation cost on every edge. 

\smallskip
\begin{standing}\label{standing:C_monotone}
	The mapping $c: \R_{\geq 0}^{m} \to \R_{\geq 0}^{e}$ is continuous and monotone.
	\hfill$\square$
\end{standing}
\smallskip
Standing Assumption~\ref{standing:C_monotone} limits the design of each $c_i(\cdot)$ to the set of continuous costs that guarantees the monotonicity of the operator obtained by stacking each $c_i(\cdot)$, $i \in \mc{E}$.
Analogously, one can define a cost on the paths as $C(\bs{p}) \coloneqq \col((C_r (\bs{p}))_{r \in \mc{M}})$, where every $C_r (\cdot)$ is typically computed as the sum of the costs on the edges generating that path, i.e., $C_r(\bs{p}) \coloneqq \sum_{i \in \mc{E}} b_{i,r} c_i(B\bs{p})$. Note that $C_r(\cdot)$ amounts to the cost experienced by each agent for choosing the path $r \in \mc{M}$. Then, the mapping $C : \R_{\geq 0}^{m} \to \R_{\geq 0}^{m}$ turns out to be
\begin{equation}\label{eq:c_cost}
	C(\bs{p}) = B^\top c(B \bs{p}).
\end{equation}

Thus, by collecting all the introduced elements, it follows that a routing game (or traffic problem) with uncertain demand is formalized as the tuple $(\mc{G}, \mc{L}, C,  \mc{P}_\omega, \Omega, \prob)$.



\subsection{Robust decisions in uncertain routing games}
The goal of an uncertain traffic problem is then to seek for some robust path flow vector $\bs{p} \in \mc{P}_\omega$, $\omega \in \Omega$, which guarantees a ``reasonable'' overall network cost (we give a rigorous definition of what ``reasonable'' means in Definition~\ref{def:user_traffic_unc}). In a fully deterministic setting, such a problem is typically formalized as a \gls{ncp} once assumed that each agent crossing the network chooses its path according to the minimum cost between every \gls{od} pair (\emph{Wardrop behavioural axiom} \cite{wardrop1952road}). The set of solutions to the \gls{ncp} happens to correspond to the set of agents' traffic equilibria, where the feasible paths that are used will have an identical cost, while the paths with costs higher than the minimum will have no flow. We formalize these concepts in the definition given next.

\smallskip
\begin{definition}\label{def:user_traffic_unc}
	A network flow $\bs{p}^\star$ is an \emph{agents' traffic equilibrium} of the traffic problem with uncertain demand $(\mc{G}, \mc{L}, C, \mc{P}_\omega, \Omega, \prob)$ if, i) $\bs{p}^\star \in \mc{P}_\omega$, for all $\omega \in \Omega$, and ii) for all \gls{od} pair $i \in \mc{L}$, and for all pair of paths $(r,s) \in \mc{M} \times \mc{M}$ connecting the $i$-th pair,
	\begin{equation}\label{eq:traffic_eq}
		C_r(\bs{p}^\star) > C_s(\bs{p}^\star) \implies p_r^\star = 0.
	\end{equation}
	\hfill$\square$
\end{definition}
\smallskip

As a static assignment problem affected by an uncertain traffic demand, it follows from \cite[Th.~3.14]{patriksson1994traffic} that, for the considered model, a network flow vector $\bs{p} \in \mc{P}_{\omega}$, $\omega \in \Omega$, satisfies the set of conditions in \eqref{eq:traffic_eq} if and only if it solves the uncertain \gls{VI} problem associated with the traffic problem $(\mc{G}, \mc{L}, C, \mc{P}_\omega, \Omega, \prob)$, denoted as VI$(\mc{P}_\omega,C)$, $\omega \in \Omega$, which coincides with the structure of uncertain \glspl{VI} analysed in \cite{paccagnan2019scenario,fabiani2020probabilistic}.
Specifically, a network flow $\bs{p}^\star \in \mc{P}_\omega$, for all $\omega \in \Omega$, amounts to an agents' traffic equilibrium if it satisfies
\begin{equation}\label{eq:unc_VI}
	(\bs{r} - \bs{p}^\star)^\top C(\bs{p}^\star) \geq 0, \, \text{ for all } \bs{r} \in \mc{P}_\omega, \, \omega \in \Omega.
\end{equation}

However, the uncertain nature of the network demand, encoded by the random parameter $\omega$ in \eqref{eq:unc_set}, drastically limits the possibilities to compute a feasible solution to such a \gls{VI}. In fact, i) the set $\Omega$ may be a-priori unknown and the only information available may come via scenarios for $\omega$; ii) even if $\Omega$ is known, it might be a set with infinite cardinality, thereby giving rise to an infinite set of constraints in \eqref{eq:unc_VI}; iii) it may happen that also the probability distribution $\prob$ can be a-priori unavailable. These reasons make the computation of an agents' traffic equilibrium prohibitive, and motivate us to investigate a data-driven approach by exploiting available realizations of the uncertain parameter $\omega$, thus looking at probabilistic feasibility certificates for traffic equilibria. 

\section{Scenario-based traffic equilibrium problems}\label{sec:scenario_traffic}
In this section, we describe the scenario-based traffic equilibrium problem associated with the routing game with uncertain demand $(\mc{G}, \mc{L}, C, \mc{P}_\omega, \Omega, \prob)$. Successively, we formalize the data-driven decision-making problem addressed.

We thus consider a $K$-multisample $\omega_K \coloneqq \{\omega^{(i)}\}_{i \in \mc{K}} = \{\omega^{(1)}, \ldots, \omega^{(K)}\} \in \Omega^K$, $\mc{K} \coloneqq \{1,2,\ldots,K\}$, as a finite collection of $K \in \N$ \gls{iid} observations of $\omega$. 
Here, every $K$-multisample is defined over the probability space $(\Omega^K, \mc{D}^K, \mathbb{P}^K)$, resulting from the $K$-fold Cartesian product of the original probability space $(\Omega,\mc{D},\mathbb{P})$.
Each sample $\omega^{(i)} \in \omega_K$ introduces a set of linear inequalities as in \eqref{eq:unc_set} described by the pair $(A(\omega^{(i)}), b(\omega^{(i)}))$, 
and we denote with $\mc{P}_{\omega_K} \coloneqq \cap_{i \in \mc{K}} \mc{P}_{\omega^{(i)}}$ the feasible set of the randomized (or scenario-based) routing game, formally defined by the tuple  $(\mc{G}, \mc{L}, C, \mc{P}_{\omega_K}, \omega_K)$.

\smallskip
\begin{standing}\label{standing:feasible_compact_nonempty}
	For any $K \in \N_0$, $\mc{P}_{\omega_K}$ is a nonempty set, for all $\omega_K \in \Omega^K$.
	\hfill$\square$
\end{standing}
\smallskip

Consequently, given any $\omega_K$, the scenario-based traffic problem $(\mc{G}, \mc{L}, C, \mc{P}_{\omega_K}, \omega_K)$ turns out to be a deterministic equilibrium problem revolving around the following notion.

\smallskip
\begin{definition}\label{def:user_traffic}
	Let $\omega_K \in \Omega^K$ be a given $K$-multisample. A network flow $\bs{p}^\star$ is an \emph{agents' traffic equilibrium} of the scenario-based traffic problem $(\mc{G}, \mc{L}, C, \mc{P}_{\omega_K}, \omega_K)$ if, i) $\bs{p}^\star \in \mc{P}_{\omega_K}$, and ii) and for all pair of paths $(r,s) \in \mc{M} \times \mc{M}$ connecting the $i$-th pair, the relation in \eqref{eq:traffic_eq} hold true.
	\hfill$\square$
\end{definition}
\smallskip

Note that, given the dependency on the $K$ realizations $\omega_K$, any agents' traffic equilibrium $\bs{p}^\star$ is a random variable. Then, according to Definition~\ref{def:user_traffic} and exploiting the scenario-based counterpart of \eqref{eq:unc_VI}, we characterize the set of traffic equilibria in terms of solution to a deterministic \gls{VI} as follows
\begin{equation}\label{eq:eq_set}
	\mc{S}_{\omega_K} \coloneqq \{\bs{p} \in \mc{P}_{\omega_K} \mid (\bs{r} - \bs{p})^\top C(\bs{p}) \geq 0, \; \forall \bs{r} \in \mc{P}_{\omega_K} \}.
\end{equation}
Specifically, after observing $K$ realizations of $\omega$, the set $\mc{S}_{\omega_K}$ contains all those network path flows that are feasible and fulfil the conditions in \eqref{eq:traffic_eq}.
For the case $K = 0$, we assume a set of nominal, deterministic constraints $\mc{P}_0$ is available to linearly bound the network demand, i.e., $\mc{P}_{\omega_0} = \mc{P}_0 \cap \mc{P}$ in \eqref{eq:unc_set}, and hence no uncertainty is present. Specifically, the set $\mc{P}_0$ is formally represented by the pair $(A_0, b_0)$, for some $A_0 \in \R^{s \times \ell}$, $b_0 \in \R^\ell$.
We characterize next the set of scenario-based agents' traffic equilibria $\mc{S}_{\omega_K}$ for a generic $K \in \N$.

\smallskip
\begin{lemma}\label{lemma:solution_compact_nonempty}
	For all $K \in \N_0$ and $\omega_K \in \Omega^K$, $\mc{S}_{\omega_K}$ is a nonempty, compact and convex set.
	\hfill$\square$
\end{lemma}
\smallskip
\begin{proof}
	See Appendix.
\end{proof}
\smallskip

Thus, given any $\omega_K$, we wish to evaluate the robustness of any agents' traffic equilibrium lying in $\mc{S}_{\omega_K}$ in \eqref{eq:eq_set} to unseen realizations of the uncertain parameter $\omega$. By denoting with $\mc{S}_{\omega} \subseteq \R^m$ the set of equilibria induced by a generic realization $\omega \in \Omega$, we introduce the following key notion.
\smallskip
\begin{definition}(\hspace{-.03mm}\textup{\cite{fabiani2020probabilistic}})\label{def:violation_set}
	The \emph{violation probability} associated to a set $\mc{S}$ is defined as
	\begin{equation}\label{eq:violation_set}
		V(\mc{S}) \coloneqq \mathbb{P}\{\omega \in \Omega \mid \mc{S} \not\subseteq \mc{S}_\omega \}.
	\end{equation}
	\hfill$\square$
\end{definition} 
\smallskip
Roughly speaking, the condition $\mc{S} \not\subseteq \mc{S}_\omega$ in \eqref{eq:violation_set} means that, once $\omega$ is drawn, at least one solution in $\mc{S}$ is lost. Therefore, the function $V : 2^{\R^m} \to [0,1]$ measures the violation of robustness of the set $\mc{S}$ to any unseen realization of $\omega$. For some reliability parameter $\varepsilon \in (0,1)$, indeed, we say that $\mc{S}$ is $\varepsilon$-robust if $V(\mc{S}) \leq \varepsilon$.  Thus, given a $K$-multisample $\omega_K \in \Omega^K$, we investigate the distribution of $V(\mc{S}_{\omega_K})$ to find a confidence bound $1-\beta$, for some given $\beta \in (0,1)$, that certifies the $\varepsilon$-robustness of $\mc{S}_{\omega_K}$,  i.e., $V(\mc{S}_{\omega_K}) \leq \varepsilon$.

\smallskip
\begin{remark}
	We do not address the computation of network traffic equilibria, as it is per se an interesting question beyond the scope of the current paper -- see, e.g., \cite{gwinner2006random,blanchini2019network,blanchini2019strategy,verbree2020stochastic,nguyen2018extragradient}.
	\hfill$\square$
\end{remark}

\section{Probabilistic feasibility certificates\\ of optimal route choice}
We recall now some key notions of the scenario approach theory that are instrumental to successively provide bounds on the violation probability related to the set of network flows of the scenario-based traffic problem $(\mc{G}, \mc{L}, C, \mc{P}_{\omega_K}, \omega_K)$.

\subsection{Scenario approach for randomized routing games}
Recent developments in the literature of the scenario theory, originally conceived to bound the out-of-sample feasibility guarantees associated with the solution to an uncertain convex optimization problem \cite{calafiore2006scenario,campi2018introduction}, provided a-posteriori probabilistic feasibility certificates for abstract decision-making problems characterized by a solution $\theta^\star_{\omega_K}$, computed after observing $K$ realizations of the uncertain parameter. In particular, \cite[Th.~1]{campi2018general} provides a distribution-free probabilistic bound $1-\beta$, for some arbitrarily fixed $\beta \in (0,1)$, which guarantees that $V(\theta^\star_{\omega_K}) \leq \varepsilon$ holds. Such a result is rooted into the two following key assumptions:
\begin{enumerate}
	\item[(i)] For all $K \in \N_0$ and all $\omega_K \in \Omega^K$, $\theta^\star_{\omega_K}$ is the \emph{unique} solution to the considered decision-making problem;
	\item[(ii)] The decision taken after observing $K$ realizations shall be \textit{consistent} with all the collected scenarios $k \in \mc{K}$.
\end{enumerate}

Since the randomized routing game $(\mc{G}, \mc{L}, C, \mc{P}_{\omega_K}, \omega_K)$ is a decision-making problem, we wish to extend the conditions above to encompass the scenario-based traffic equilibrium problem described in \S \ref{sec:scenario_traffic}, and thereby apply the distribution-free, probabilistic feasibility bound in \cite[Th.~1]{campi2018general} to any given agents' traffic equilibrium within the set $\mc{S}_{\omega_K}$. 
Along this direction, the traditional scenario approach has been recently investigated from a set-oriented perspective. In particular, by focusing on a family of uncertain \glspl{VI} as specific class of decision-making problems, the works in \cite{fabiani2020scenario,fabiani2020probabilistic} extended \cite[Th.~1]{campi2018general} to a set-oriented framework. As a main result, they provided a-posteriori robustness certificates for the entire set of solutions to uncertain \glspl{VI} of the form VI$(\mc{P}_\omega,C)$, $\omega \in \Omega$. Therefore, given the variational nature of the considered scenario-based traffic problem $(\mc{G}, \mc{L}, C, \mc{P}_{\omega_K}, \omega_K)$, and by letting coincides our decision to the entire set of equilibria, $\mc{S}_{\omega_K}$, we aim to pave the way for applying such certificates, and, specifically, \cite[Th.~1]{fabiani2020probabilistic}, to the route choice in traffic problems. 
We briefly discuss next the key points that allow one to apply the scenario theory to a set-oriented framework, thus emphasizing how to translate items \textrm{(i)}--\textrm{(ii)} above to set of traffic equilibria $\mc{S}_{\omega_K}$ for $(\mc{G}, \mc{L}, C, \mc{P}_{\omega_K}, \omega_K)$.

\subsubsection{Uniqueness of the set of decisions}
We define $\Theta_K : \Omega^K \rightrightarrows \mc{P}$ as the mapping that, given a set of realizations $\omega_K$, returns the set of agents' traffic equilibria, i.e.,
\begin{equation}\label{eq:theta_mapping}
	\Theta_K(\omega^{(1)}, \ldots, \omega^{(K)}) = \Theta_K(\omega_K) \coloneqq \mc{S}_{\omega_K} .
\end{equation}
When $K = 0$, we assume that $\Theta_0$ returns the set of equilibria $\mc{S}_{\omega_0}$.
In view of item \textrm{(i)}, the uniqueness of the solution returned by $\Theta_K$ holds by definition since, for any $\omega_K \in \Omega^K$, there is naturally a single set of traffic equilibria that is nonempty, compact and convex \cite[Th.~2.3.5]{facchinei2007finite}. 

\subsubsection{Consistency of the set of decisions} To address item \textrm{(ii)}, which essentially coincides with \cite[Ass.~1]{campi2018general}, we introduce a consistency property for the set of agents' traffic equilibria $\mc{S}_{\omega_K}$ as follows.
\smallskip
\begin{definition}(\hspace{-.03mm}\textup{\cite[Def.~4]{fabiani2020probabilistic}})\label{def:cons_set}
	Given any $K \in \N$ and $\omega_K \in \Omega^K$, the set of traffic equilibria of the scenario-based traffic equilibrium problem $(\mc{G}, \mc{L}, C, \mc{P}_{\omega_K}, \omega_K)$ is \emph{consistent} with the collected scenarios if $\Theta_K(\omega_K) = \mc{S}_{\omega_K} \subseteq \mc{P}_{\omega^{(i)}}$, for all $i \in \mc{K}$.
	\hfill$\square$
\end{definition}
\smallskip
Specifically, Definition~\ref{def:cons_set} establishes that the set of agents' traffic equilibria, $\mc{S}_{\omega_K}$, which is based on $K$ scenarios, should be feasible for each of the sets $\mc{P}_{\omega^{(i)}}$, $i \in \mc{K}$, corresponding to each of the $K$ realizations of the uncertain parameter. 
Note that, for any $K \in \N$ and associated $\omega_{K} \in \Omega^K$, we have $\Theta_K(\omega_K) \coloneqq \mc{S}_{\omega_K} \subseteq \cap_{i \in \mc{K}} \mc{P}_{\omega^{(i)}}$, which on the other hand implies that $\Theta_K(\omega_K) \subseteq \mc{P}_{\omega^{(i)}}$, for all $i \in \mc{K}$, thus falling within Definition~\ref{def:cons_set}.
It therefore holds by definition that the mapping $\Theta_K(\cdot)$ is consistent with the realizations observed in the randomized routing game $(\mc{G}, \mc{L}, C, \mc{P}_{\omega_K}, \omega_K)$. 

Now, given some $K \in \N$, let $\mc{S}_{\omega_{K+1}} \coloneqq \mc{S}_{\omega_K \cup \{\omega^{(K+1)}\}}$ be the set of agents' traffic equilibria to the randomized routing game $(\mc{G}, \mc{L}, C, \mc{P}_{\omega_{K+1}}, \omega_K \cup \{\omega^{(K+1)}\})$ after observing the $(K+1)$-th realization of $\omega$. We show next a result that links the sets $\mc{S}_{\omega_{K}}$ and $\mc{S}_{\omega_{K+1}}$ across the samples scenarios, thus establishing the set-oriented counterpart of \cite[Ass.~1]{campi2018general}.
\smallskip
\begin{lemma}\label{lemma:if_affine}
	If for all $K \in \N_0$ and $\omega_K \in \Omega^K$, $\mathrm{aff}(\mc{S}_{\omega_K}) = \mathrm{aff}(\mc{S}_{\omega_0})$, then it holds that $\mc{S}_{\omega_{K+1}} = \mc{S}_{\omega_K} \cap \mc{P}_{\omega^{(K+1)}}$.
	\hfill$\square$
\end{lemma}
\smallskip
\begin{proof}
	See Appendix.
\end{proof}
\smallskip

\begin{figure}[t!]
	\centering
	\includegraphics[width=.9\columnwidth]{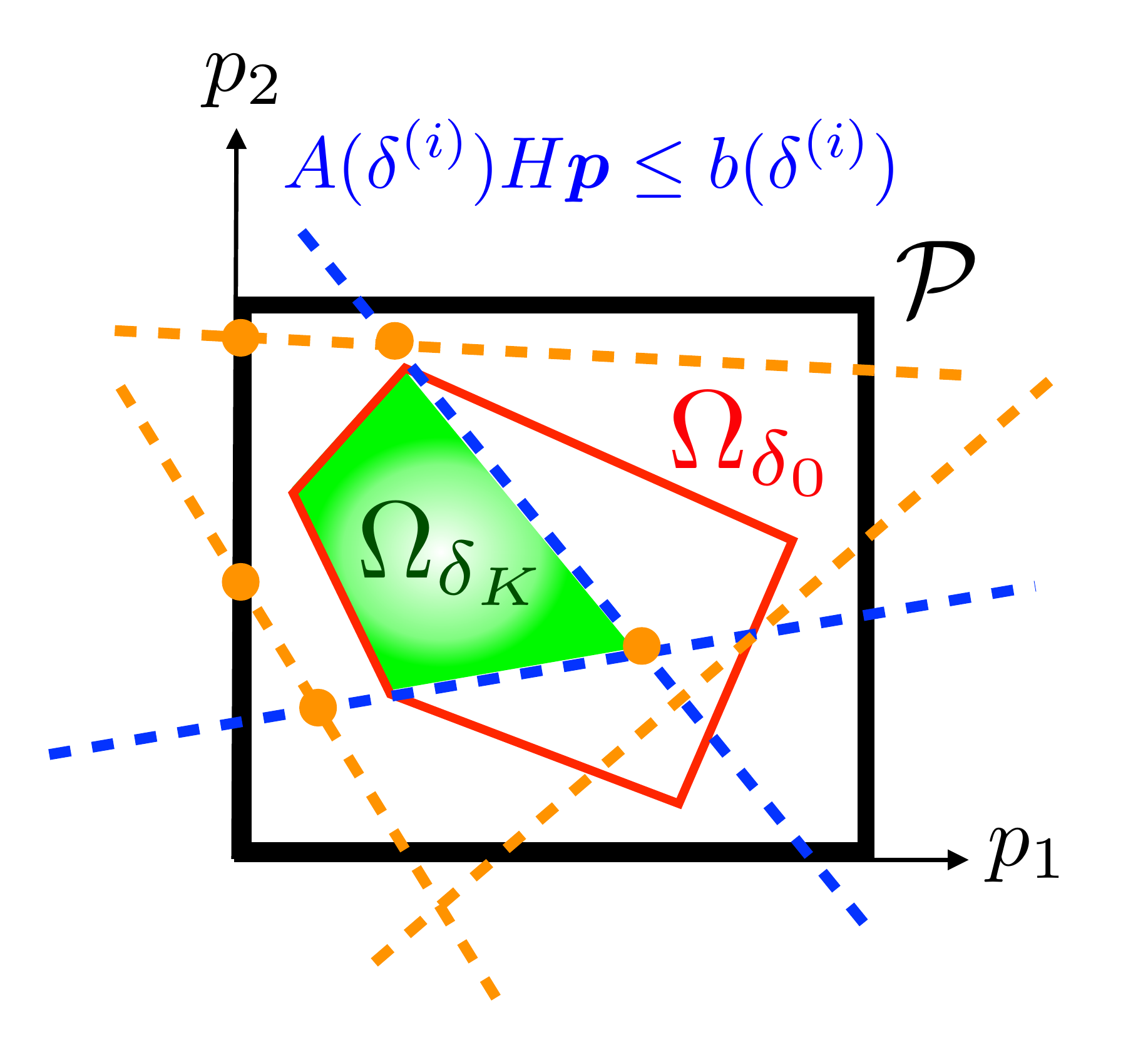}
	\caption{Schematic two dimensional interpretation of Lemma~\ref{lemma:if_affine}. In this case, $\mc{P}_{\omega_{K}}$ coincides with the convex hull of the orange dots, while the set of agent's traffic equilibria $\mc{S}_{\omega_{K}}$ (green region) associated to the randomized routing game $(\mc{G}, \mc{L}, C, \mc{P}_{\omega_K}, \omega_K)$
can be ``shaped" by the set of linear constraints modelling uncertain traffic demand configurations (dashed blue lines). According to Definition~\ref{def:support_sub}, the dashed orange lines still denote linear constraints modelling uncertain traffic demands, which however are not of support. }
	\label{fig:shape_set}
\end{figure}

The assumption in Lemma~\ref{lemma:if_affine} essentially rules out the scenario that an observed realization of the $K$-multisample $\omega_K$ reduces the set of agents' traffic equilibria $\mc{S}_{\omega_K}$ to one of lower dimension compared to the set collecting those flows in $\mc{P}_{\omega_0}$ that satisfy the conditions in \eqref{eq:traffic_eq}, i.e., $\mc{S}_{\omega_0}$. Specifically, it has been identified as one of the
weakest conditions allowing proof of consistency of a set $\mc{S}_{\omega_K}$ by
relying on known results available in the literature. Following the discussion in \cite[\S 8]{campi2018wait}, it appears that the non-degeneracy assumption postulated in similar works (e.g., \cite{campi2018wait,paccagnan2019scenario}) is
definitely not easier to verify than, for all $K \in \N_0$ and $\omega_K \in \Omega^K$, $\mathrm{aff}(\mc{S}_{\omega_K}) = \mathrm{aff}(\mc{S}_{\omega_0})$. Additionally, this latter has a clear geometrical interpretation reflecting
onto the probability space $\Omega$ (see \cite[\S 3.2]{fabiani2020probabilistic} for additional details).
The pictorial interpretation of Lemma~\ref{lemma:if_affine} in Fig.~\ref{fig:shape_set} shows that generating samples gives rise to uncertain traffic polyhedra that “shape” the set of agents' traffic equilibria $\mc{S}_{\omega_K}$. As an immediate consequence, we have that the inclusions $\Theta_0 \coloneqq \mc{S}_{\omega_0} \supseteq \mc{S}_{\omega_1} \supseteq \ldots \supseteq \mc{S}_{\omega_K} \eqqcolon \Theta_K(\omega_K)$ intrinsically hold.

\subsection{Probabilistic feasibility guarantees for traffic equilibria}
Before stating the main result, we recall the following definition from  \cite[Def.~2]{campi2018general}, which is at the core of the scenario approach theory and our subsequent derivation.
\smallskip
\begin{definition}\label{def:support_sub}
	Given any $K \in \N$ and associated $K$-multisample, $\omega_K \in \Omega^K$, a \emph{support subsample} $\{\omega^{(i_1)}, \ldots, \omega^{(i_p)}\} \subseteq \omega_K$ is a $p$-tuple of unique elements of $\omega_K$, $i_1 < \ldots < i_p$, that satisfies
	$
	\Theta_p(\omega^{(i_1)}, \ldots, \omega^{(i_p)}) = \Theta_K(\omega^{(1)}, \ldots, \omega^{(K)}),
	$
	i.e., it gives the same set of agents' traffic equilibria as the original sample.
	\hfill$\square$
\end{definition}
\smallskip

Then, let $\Upsilon_K : \Omega^K \rightrightarrows \mc{K}$ be any algorithm returning a $p$-tuple $\{i_1, \ldots, i_p\}$, $i_1 < \ldots < i_p$,  such that $\{\omega^{(i_1)}, \ldots, \omega^{(i_p)}\}$ is a support subsample for $\omega_K$, and let $\iota_K \coloneqq |\Upsilon_K(\omega_K)|$. With the postulated assumptions, examples of efficient algorithms $\Upsilon_K(\cdot)$ that accomplish this task are represented by \cite[Alg.~1]{fabiani2020probabilistic} and \cite[\S II]{campi2018general}.
Note that $\iota_K$ is itself a random variable, as it depends on $\omega_K$. 
Our main result, which follows directly from \cite{fabiani2020probabilistic}, characterizes the violation probability of $\mc{S}_{\omega_K}$, i.e., any agents' traffic equilibrium of the randomized routing game $(\mc{G}, \mc{L}, C, \mc{P}_{\omega_{K}}, \omega_K)$, as stated next.
\smallskip
\begin{proposition}\label{prop:VI}
	Fix $\beta \in (0,1)$, and let $\varepsilon : \mc{K} \cup \{0\} \to [0, 1]$ be a function such that
	\begin{equation}\label{eq:epsilon}
		\left\{\begin{aligned}
			& \varepsilon(K) = 1,\\
			& \sum_{h = 0}^{K - 1} \left( \begin{array}{c}
				K\\
				h
			\end{array} \right) (1 - \varepsilon(h))^{K - h} = \beta.
		\end{aligned}
		\right.
	\end{equation}
	If for all $K \in \N_0$ and $\omega_K \in \Omega^K$, $\mathrm{aff}(\mc{S}_{\omega_K}) = \mathrm{aff}(\mc{S}_{\omega_0})$, then for any mappings $\Theta_K$, $\Upsilon_K$ and distribution $\mathbb{P}$, it holds that
	\begin{equation}\label{eq:prob_feas_boud}
		\mathbb{P}^K \{\omega_K \in \Omega^K \mid V(\mc{S}_{\omega_K}) > \varepsilon(\iota_K) \} \leq \beta.
	\end{equation}
	\hfill$\square$
\end{proposition}
\smallskip
\begin{proof}
	The proof follows immediately by noting that the scenario-based traffic problem $(\mc{G}, \mc{L}, C, \mc{P}_{\omega_{K}}, \omega_K)$ satisfies all the assumptions required by \cite[Th.~1]{fabiani2020probabilistic}.
\end{proof}
\smallskip

Some considerations following Proposition~\ref{prop:VI} are in order. First, we remark that Proposition~\ref{prop:VI} is a distribution-free result, i.e., one does not need to know $\prob$ to rely on the bound in \eqref{eq:prob_feas_boud}. Essentially, it implies that the probability that $\mc{S}_{\omega_K \cup \{\omega\}}$ differs from $\mc{S}_{\omega_K}$ is at most equal to $\varepsilon(\iota_K)$, with confidence at least $1-\beta$, for an arbitrarily small $\beta \in (0,1)$. In addition, we remark that the probabilistic bound in \eqref{eq:prob_feas_boud} is an a-posteriori statement, as $\iota_K$ depends on the observed $K$-multisample $\omega_K$. Finally, as evident from \eqref{eq:prob_feas_boud}, to certify the robustness of any agents' traffic equilibrium in $\mc{S}_{\omega_K}$, one does not need a full characterization of $\mc{S}_{\omega_K}$ itself, but rather the number of support subsamples, according to Definition~\ref{def:support_sub}. 

\section{Numerical example}
\begin{figure}[t!]
	\centering
	\includegraphics[width=1.0\columnwidth]{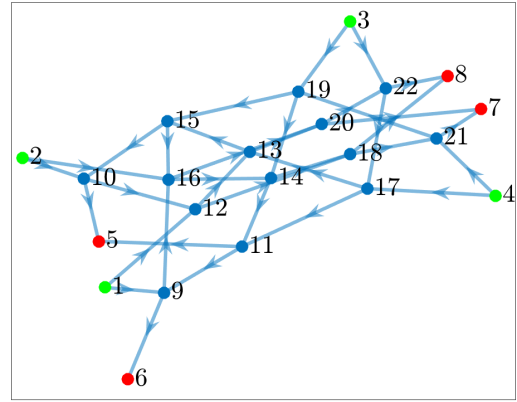}
	\caption{Traffic network digraph. The green dots denote the origin nodes, while the red dots the destination ones.}
	\label{fig:graph_top}
\end{figure}
We support our theoretical findings with a numerical example borrowed from \cite{nagurney1984comparative} and, specifically, Network~$26$. The simulations are run in Matlab on a laptop with a Quad-Core Intel Core i5 2.4 GHz CPU and 8 Gb RAM. As depicted in Fig.~\ref{fig:graph_top}, we consider a traffic network digraph consisting of $22$ nodes and $36$ edges, with \gls{od} pairs specified in $\mc{L} = \{(1,6), (1,7), (1,8), (2,5), (2,7), (2,8), (3,5), (3,6), (3,8),$ $ (4,5), (4,6), (4,7)\}$, and hence $\ell = 12$. By enumerating all possible paths connecting each \gls{od} pair in $\mc{L}$, it turns out that the considered traffic network is characterized by a total of $m = 124$ path flows with $\bs{p}_{\textrm{max}} = 50 \times \bsone_{124}$. 
Moreover, the pair $(A_0, b_0)$, which describes the deterministic, nominal set $\mc{P}_0$, coincides with the hyperplane representation of the polytope originated by the convex hull of $15$ random points sampled in $H\mc{P} =  [\bs{0}, H \bs{p}_{\textrm{max}}]$, thus resulting in $s = 136$ linear inequalities. Starting from the nominal pair $(A_0, b_0)$, we assume the uncertainty affects additively the vector $b(\cdot)$ only, i.e., $b(\omega) = b_0 + \omega$ and $A(\omega) = A_0$, for all $\omega \in \Omega = b_0 \times [-0.5, 0.5] \subseteq \R^{136}$, with $\omega$ following a uniform probability distribution on $\Omega$. Note that the uncertainty set supports modelling errors in the nominal ``offset''  $b_0$ of the linear constraints up to the $50 \%$.
\begin{figure}
	\centering
	\includegraphics[width=1.0\columnwidth]{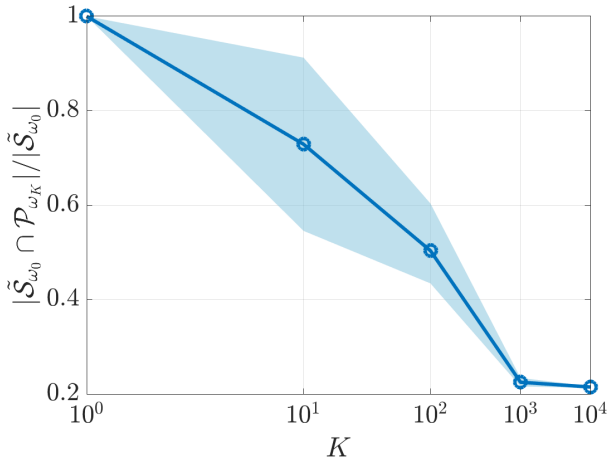}
	\caption{Normalized cardinality of the set $\tilde{\mc{S}}_{\omega_K} $, averaged over $10$ numerical experiments, as a function of the samples $K$ (solid line). The shaded area denotes the standard deviation.}\label{fig:reduction}
\end{figure}

\begin{figure}[t!]
	\centering
	\includegraphics[width=1.0\columnwidth]{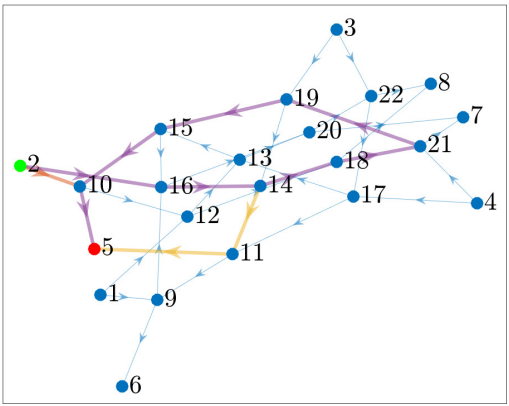}
	\caption{Path flows connecting the \gls{od} pair $(2,5)$. According to Definition~\ref{def:user_traffic}, out of a total of $13$ paths, only three of them have nonnull flows (violet, yellow and red edges).}
	\label{fig:graph_path}
\end{figure}

\begin{table}[t!]
	\caption{Path flows connecting the \gls{od} pair $(2,5)$.}
	\label{tab:path_flows}
	\begin{center}
		\begin{tabular}{p{20pt} p{160pt} p{20pt}}
			\toprule
			Path \# & Node sequence ($2 \rightarrow ...  \rightarrow 5$) & Value\\
			\midrule
			$p_{30}$ & \tiny{$10$} & $50$\\
			$p_{31}$ & \tiny{$10 \rightarrow 12 \rightarrow 13 \rightarrow 15 \rightarrow 16 \rightarrow 14 \rightarrow 11$} & $0$\\
			$p_{32}$ & \tiny{$10 \rightarrow 12 \rightarrow 13 \rightarrow 15 \rightarrow 16 \rightarrow 20 \rightarrow 22 \rightarrow 17 \rightarrow 11$} & $0$\\
			$p_{33}$ & \tiny{$10 \rightarrow 12 \rightarrow 13 \rightarrow 20 \rightarrow 22 \rightarrow 17 \rightarrow 11$} & $0$\\
			$p_{34}$ & \tiny{$10 \rightarrow 12 \rightarrow 18 \rightarrow 21 \rightarrow 19 \rightarrow 14 \rightarrow 11$} & $0$\\
			$p_{35}$ & \tiny{$10 \rightarrow 12 \rightarrow 18 \rightarrow 21 \rightarrow 19 \rightarrow 15 \rightarrow 16 \rightarrow 14 \rightarrow 11$} & $0$\\
			$p_{36}$ & \tiny{$10 \rightarrow 12 \rightarrow 18 \rightarrow 21 \rightarrow 19 \rightarrow 15 \rightarrow 16 \rightarrow 20 \rightarrow 22 \rightarrow 17 \rightarrow 11$} & $0$\\
			$p_{37}$ & \tiny{$16 \rightarrow 14 \rightarrow 11$} & $50$\\
			$p_{38}$ & \tiny{$16 \rightarrow 14 \rightarrow 18 \rightarrow 21 \rightarrow 19 \rightarrow 15 \rightarrow 10$} & $36.32$\\
			$p_{39}$ & \tiny{$16 \rightarrow 14 \rightarrow 18 \rightarrow 21 \rightarrow 19 \rightarrow 15 \rightarrow 10 \rightarrow 12 \rightarrow 13 \rightarrow 20 \rightarrow 22 \rightarrow 17 \rightarrow 11$} & $0$\\
			$p_{40}$ & \tiny{$16 \rightarrow 20 \rightarrow 22 \rightarrow 17 \rightarrow 11$} & $0$\\
			$p_{41}$ & \tiny{$16 \rightarrow 20 \rightarrow 22 \rightarrow 17 \rightarrow 13 \rightarrow 15 \rightarrow 10$} & $0$\\
			$p_{42}$ & \tiny{$16 \rightarrow 20 \rightarrow 22 \rightarrow 17 \rightarrow 13 \rightarrow 15 \rightarrow 10 \rightarrow 12 \rightarrow 18 \rightarrow 21 \rightarrow 19 \rightarrow 14 \rightarrow 11$} & $0$\\
			\bottomrule
		\end{tabular}
	\end{center}
\end{table}

Then, to numerically corroborate Lemma~\ref{lemma:if_affine}, we run an extragradient algorithm \cite{nguyen2018extragradient} with fixed step-size $\alpha = 0.3$ from $10^4$ initial points, randomly sampled in $\mc{P}_0$, to estimate the ``nominal'' set of agents' traffic equilibria $\mc{S}_{\omega_0}$, thus obtaining $\tilde{\mc{S}}_{\omega_0}$.
As illustrated in Fig.~\ref{fig:reduction}, the average number of traffic equilibria for the randomized routing game gathered in $\tilde{\mc{S}}_{\omega_K}$ over $10$ numerical experiments, normalized w.r.t. $\tilde{\mc{S}}_{\omega_0}$, shrinks as $K$ grows. 
In addition, as the number of samples $K$ increases, the standard deviation of $\omega$ narrows around the average: this fact is mainly due because of the structure of the support set $\Omega$, as well as of the type of uncertainty we consider. An example of path flows connecting the \gls{od} pair $(2,5)$ is shown in Fig.~\ref{fig:graph_path} and Table~\ref{tab:path_flows} where, according to the Wardrop behavioural axiom and Definition~\ref{def:user_traffic}, only few paths have non-zero flow, i.e., those ones guaranteeing an overall minimal cost, while the remaining paths have no flow.

\begin{table}[t!]
	\caption{Robustness certificate \eqref{eq:prob_feas_boud} and empirical violation probability of the set of agents' traffic equilibria for the randomized routing game $(\mc{G}, \mc{L}, C, \mc{P}_{\omega_{K}}, \omega_K)$.}
 	\label{tab:feas_bound}
	\begin{center}
		\begin{tabular}{p{20pt} p{20pt} p{35pt} p{55pt} p{55pt}}
			\toprule
			$K$ & $\iota^\star_K$  & $\varepsilon(\iota^\star_K)$ & $V_{\textrm{max}}(\tilde{\mc{S}}_{\omega_K})$ & $\mathrm{avg}(V_{\textrm{max}}(\tilde{\mc{S}}_{\omega_K}))$\\
			\midrule
			$10^2$ & $8$ & $0.38$ & $21\times10^{-3}$ & $16\times10^{-3}$\\
			$10^3$ & $19$ & $0.10$ & $6.7\times10^{-3}$ & $4.2\times10^{-3}$\\
			$10^4$ & $27$ & $20\times10^{-3}$ & $1.4\times10^{-3}$ & $0.7\times10^{-3}$\\
			\bottomrule
		\end{tabular}
	\end{center}
\end{table}

Next, we obtain the analytic expression of the function $\varepsilon(\cdot)$ by splitting $\beta = 10^{-6}$ evenly among the $K$ terms within the summation in \eqref{eq:epsilon}. In this way, by adopting \cite[Alg.~1]{fabiani2020probabilistic} to enumerate the support subsamples w.r.t. $\mc{S}_{\omega_K}$, Table~\ref{tab:feas_bound} compares the maximum and the average value of the empirical violation probability of the traffic equilibria $\tilde{\mc{S}}_{\omega_0} \cap \mc{P}_{\omega_K}$ computed against $10^2$, $10^3$ and $10^4$ unobserved samples.
The empirical violation probability, as expected, is lower than the theoretical bound in Proposition~\ref{prop:VI}.
\section{Conclusion}
In the realm of uncertain routing games, making a-priori traffic predictions on the basis of available data may not be trivial, and therefore assuming some kind of uncertainty associated to the traffic demand configurations provides a key degree of freedom for the practitioners. In this context, the scenario approach paradigm enables one to assess the robustness properties of the entire set of agents' traffic equilibria. We have shown that the proposed certificates merely require one to enumerate the active constraints that intersect such set, without requiring an explicit characterization of it. 

\appendix
\section{Proofs}
\emph{Proof of Lemma~\ref{lemma:solution_compact_nonempty}:} 
	First, we note that $\mc{P}_{\omega_K}$ is a nonempty, compact and convex set. Specifically, for any $K \in \N_0$ and $\omega_K \in \Omega^K$, nonemptiness follows in view of Standing Assumption~\ref{standing:feasible_compact_nonempty}, while compactness and convexity from the fact that, according to \eqref{eq:unc_set}, $\mc{P}_{\omega_K}$ is given as the finite intersection between a set of nonnegative box-constraints lying in the positive orthant of $\R^m$, $\mc{P}$, which is compact and convex, and a collection of linear inequalities $A(\omega^{(i)}) H \, \bs{p} \leq b(\omega^{(i)})$, $i \in \mc{K}$, which are closed and convex as well \cite{brondsted2012introduction}.
Moreover, we note that the mapping $C(\cdot)$ is continuous in view of its own definition in \eqref{eq:c_cost}, as it is a positive combination (see \eqref{eq:B_def}) of the nonnegative elements of the continuous mapping $c(\cdot)$ (Standing Assumption~\ref{standing:C_monotone}). In addition, $C(\cdot)$ is also a monotone mapping since, for all $\bs{p}$, $\bs{r} \in \R_{\geq 0}^m$, it holds that:
$$
\begin{aligned}
	(C(\bs{p}) -  C(\bs{r}))^\top (\bs{p} - \bs{r}) &= (B^\top (c(B \bs{p}) - c(B \bs{r})))^\top (\bs{p} - \bs{r})\\
	&= (c(B \bs{p}) - c(B \bs{r}))^\top B (\bs{p} - \bs{r})\\
	& = (c(B \bs{p}) - c(B \bs{r}))^\top (B\bs{p} - B\bs{r})\\
	& \geq 0,
\end{aligned}
$$
where the last inequality is entailed by the monotonicity of $c(\cdot)$ (Standing Assumption~\ref{standing:C_monotone}).
Then, the statement follows by combining the results in \cite[Cor.~2.2.5, Th.~2.3.5]{facchinei2007finite}.
\hfill $\blacksquare$

\smallskip

\emph{Proof of Lemma~\ref{lemma:if_affine} (sketch):} Once proved that, for all $K \in \N_0$ and associated $K$-multisample $\omega_K$, the feasible set $\mc{P}_{\omega_K}$ is a nonempty, compact and convex set (Lemma~\ref{lemma:solution_compact_nonempty}), with the condition in the statement of Lemma~\ref{lemma:if_affine} the proof is a verbatim copy of that of \cite[Lemma~4]{fabiani2020probabilistic}. However, for the sake of completeness, we give a simplified sketch below.

Since the uncertain parameter enters in \eqref{eq:unc_set} only, and hence it does not affect the mapping $C(\cdot)$, in view of Lemma~\ref{lemma:solution_compact_nonempty} and of the convexity and compactness of the sets involved, the inclusion $\mc{S}_{\omega_K} \cap \mc{P}_{\omega^{(K+1)}} \subseteq \mc{S}_{\omega_{K+1}}$ follows immediately. 

To show the reverse inclusion, i.e., $\mc{S}_{\omega_K} \cap \mc{P}_{\omega^{(K+1)}} \supseteq \mc{S}_{\omega_{K+1}}$, the condition in the statement of Lemma~\ref{lemma:if_affine} allows us to resort \cite[Cor.~1.6.1]{rockafellar1970convex} to treat $\mc{S}_{\omega_{K+1}}$, and hence $\mc{S}_{\omega_K}$, as an $m$-dimensional set, i.e., such that $\mathrm{relint}(\mc{S}_{\omega_{K+1}}) = \mathrm{int}(\mc{S}_{\omega_{K+1}}) \neq \emptyset$. First of all, we can exclude the case in which there exists some $\bs{p}^\star \in \mc{P}_{\omega_K} \cap \mc{P}_{\omega^{(K+1)}}$ such that $\bs{p}^\star \in \mathrm{int}(\mc{S}_{\omega_{K+1}})$, but $\bs{p}^\star \notin \mc{S}_{\omega_K}$. This follows as a consequence of the definition of the normal cone, as $\bs{p}^\star \in \mathrm{int}(\mc{S}_{\omega_{K+1}}) \subseteq \mathrm{int}(\mc{P}_{\omega_K} \cap \mc{P}_{\omega^{(K+1)}}) \subseteq \mathrm{int}(\mc{P}_{\omega_K})$ if and only if $-C(\bs{p}^\star) \in \mc{N}_{\mc{P}_{\omega_{K+1}}}(\bs{p}^\star) = \{\bs{0}\}$. Finally, in case $\bs{p}^\star \in \mathrm{bdry}(\mc{S}_{\omega_{K+1}})$, since $\mathrm{relint}(\mc{S}_{\omega_{K+1}}) \neq \emptyset$, it follows from \cite[Th.~6.1]{rockafellar1970convex} that we can always construct a convergent sequence of points $\{\bs{p}_t\}_{t \in \N}$ such that, for all $t \in \N$, $\bs{p}_t \in \mathrm{relint}(\mc{S}_{\omega_{K+1}}) \subseteq \mc{S}_{\omega_K}$, and $\{\bs{p}_t\}_{t \in \N} \to \bs{p}^\star$, implying that $\bs{p}^\star \in \mc{S}_{\omega_K}$.
\hfill $\blacksquare$


\balance
\bibliographystyle{IEEEtran}
\bibliography{21_CDC_unc_traffic}

\begin{thebibliography}{10}
\providecommand{\url}[1]{#1}
\csname url@samestyle\endcsname
\providecommand{\newblock}{\relax}
\providecommand{\bibinfo}[2]{#2}
\providecommand{\BIBentrySTDinterwordspacing}{\spaceskip=0pt\relax}
\providecommand{\BIBentryALTinterwordstretchfactor}{4}
\providecommand{\BIBentryALTinterwordspacing}{\spaceskip=\fontdimen2\font plus
\BIBentryALTinterwordstretchfactor\fontdimen3\font minus
  \fontdimen4\font\relax}
\providecommand{\BIBforeignlanguage}[2]{{%
\expandafter\ifx\csname l@#1\endcsname\relax
\typeout{** WARNING: IEEEtran.bst: No hyphenation pattern has been}%
\typeout{** loaded for the language `#1'. Using the pattern for}%
\typeout{** the default language instead.}%
\else
\language=\csname l@#1\endcsname
\fi
#2}}
\providecommand{\BIBdecl}{\relax}
\BIBdecl

\bibitem{smith1979existence}
M.~J. Smith, ``The existence, uniqueness and stability of traffic equilibria,''
  \emph{Transportation Research Part B: Methodological}, vol.~13, no.~4, pp.
  295--304, 1979.

\bibitem{dafermos1980traffic}
S.~Dafermos, ``Traffic equilibrium and variational inequalities,''
  \emph{Transportation Science}, vol.~14, no.~1, pp. 42--54, 1980.

\bibitem{florian1995network}
M.~Florian and D.~Hearn, ``Network equilibrium models and algorithms,''
  \emph{Handbooks in Operations Research and Management Science}, vol.~8, pp.
  485--550, 1995.

\bibitem{patriksson1994traffic}
M.~Patriksson, \emph{The traffic assignment problem: models and methods}, ser.
  Topics in Transportation Series.\hskip 1em plus 0.5em minus 0.4em\relax VSP,
  1994.

\bibitem{gojmerac2003adaptive}
I.~Gojmerac, T.~Ziegler, F.~Ricciato, and P.~Reichl, ``Adaptive multipath
  routing for dynamic traffic engineering,'' in \emph{GLOBECOM'03. IEEE Global
  Telecommunications Conference (IEEE Cat. No. 03CH37489)}, vol.~6.\hskip 1em
  plus 0.5em minus 0.4em\relax IEEE, 2003, pp. 3058--3062.

\bibitem{chiang2007layering}
M.~Chiang, S.~H. Low, A.~R. Calderbank, and J.~C. Doyle, ``Layering as
  optimization decomposition: A mathematical theory of network architectures,''
  \emph{Proceedings of the IEEE}, vol.~95, no.~1, pp. 255--312, 2007.

\bibitem{wardrop1952road}
J.~G. Wardrop, ``Road paper. {S}ome theoretical aspects of road traffic
  research.'' \emph{Proceedings of the Institution of Civil Engineers}, vol.~1,
  no.~3, pp. 325--362, 1952.

\bibitem{fingerhut1997designing}
J.~A. Fingerhut, S.~Suri, and J.~S. Turner, ``Designing least-cost nonblocking
  broadband networks,'' \emph{Journal of Algorithms}, vol.~24, no.~2, pp.
  287--309, 1997.

\bibitem{ben2005routing}
W.~Ben-Ameur and H.~Kerivin, ``Routing of uncertain traffic demands,''
  \emph{Optimization and Engineering}, vol.~6, no.~3, pp. 283--313, 2005.

\bibitem{gwinner2006random}
J.~Gwinner and F.~Raciti, ``Random equilibrium problems on networks,''
  \emph{Mathematical and Computer Modelling}, vol.~43, no. 7-8, pp. 880--891,
  2006.

\bibitem{ouorou2007model}
A.~Ouorou and J.-P. Vial, ``A model for robust capacity planning for
  telecommunications networks under demand uncertainty,'' in \emph{2007 6th
  International workshop on design and reliable communication networks}.\hskip
  1em plus 0.5em minus 0.4em\relax IEEE, 2007, pp. 1--4.

\bibitem{lemarechal2010robust}
C.~Lemar{\'e}chal, A.~Ouorou, and G.~Petrou, ``Robust network design in
  telecommunications under polytope demand uncertainty,'' \emph{European
  Journal of Operational Research}, vol. 206, no.~3, pp. 634--641, 2010.

\bibitem{frangioni2011static}
A.~Frangioni, F.~Pascali, and M.~G. Scutell{\`a}, ``Static and dynamic routing
  under disjoint dominant extreme demands,'' \emph{Operations Research
  Letters}, vol.~39, no.~1, pp. 36--39, 2011.

\bibitem{ouorou2013tractable}
A.~Ouorou, ``Tractable approximations to a robust capacity assignment model in
  telecommunications under demand uncertainty,'' \emph{Computers \& Operations
  Research}, vol.~40, no.~1, pp. 318--327, 2013.

\bibitem{daniele2015random}
P.~Daniele and S.~Giuffr{\`e}, ``Random variational inequalities and the random
  traffic equilibrium problem,'' \emph{Journal of Optimization Theory and
  Applications}, vol. 167, no.~1, pp. 363--381, 2015.

\bibitem{cominetti2015equilibrium}
R.~Cominetti, ``Equilibrium routing under uncertainty,'' \emph{Mathematical
  Programming}, vol. 151, no.~1, pp. 117--151, 2015.

\bibitem{jadamba2018efficiency}
B.~Jadamba, M.~Pappalardo, and F.~Raciti, ``Efficiency and vulnerability
  analysis for congested networks with random data,'' \emph{Journal of
  Optimization Theory and Applications}, vol. 177, no.~2, pp. 563--583, 2018.

\bibitem{cherukuri2019sample}
A.~Cherukuri, ``Sample average approximation of {CVaR}-based {W}ardrop
  equilibrium in routing under uncertain costs,'' in \emph{2019 IEEE 58th
  Conference on Decision and Control (CDC)}.\hskip 1em plus 0.5em minus
  0.4em\relax IEEE, 2019, pp. 3164--3169.

\bibitem{paccagnan2019scenario}
D.~Paccagnan and M.~C. Campi, ``The scenario approach meets uncertain game
  theory and variational inequalities,'' in \emph{2019 IEEE 58th Conference on
  Decision and Control (CDC)}.\hskip 1em plus 0.5em minus 0.4em\relax IEEE,
  2019, pp. 6124--6129.

\bibitem{fabiani2020probabilistic}
F.~Fabiani, K.~Margellos, and P.~J. Goulart, ``Probabilistic feasibility
  guarantees for solution sets to uncertain variational inequalities,''
  \emph{Automatica}, 2021, (Under review -- available at {\footnotesize
  \texttt{https://arxiv.org/abs/2005.09420}}).

\bibitem{mesbahi2010graph}
M.~Mesbahi and M.~Egerstedt, \emph{Graph theoretic methods in multiagent
  networks}.\hskip 1em plus 0.5em minus 0.4em\relax Princeton University Press,
  2010, vol.~33.

\bibitem{facchinei2007finite}
F.~Facchinei and J.~S. Pang, \emph{Finite-dimensional variational inequalities
  and complementarity problems}.\hskip 1em plus 0.5em minus 0.4em\relax
  Springer Science \& Business Media, 2007.

\bibitem{blanchini2019network}
F.~Blanchini, D.~Casagrande, F.~Fabiani, G.~Giordano, and R.~Pesenti,
  ``Network-decentralised optimisation and control: An explicit saturated
  solution,'' \emph{Automatica}, vol. 103, pp. 379--389, 2019.

\bibitem{blanchini2019strategy}
------, ``A network-decentralised strategy for shortest-path-flow routing,'' in
  \emph{2019 IEEE 58th Conference on Decision and Control (CDC)}.\hskip 1em
  plus 0.5em minus 0.4em\relax IEEE, 2019, pp. 1126--1131.

\bibitem{verbree2020stochastic}
J.~Verbree and A.~Cherukuri, ``Stochastic approximation for {CVaR}-based
  variational inequalities,'' in \emph{2020 59th IEEE Conference on Decision
  and Control (CDC)}.\hskip 1em plus 0.5em minus 0.4em\relax IEEE, 2020, pp.
  2216--2221.

\bibitem{nguyen2018extragradient}
T.~P. Nguyen, E.~Pauwels, E.~Richard, and B.~W. Suter, ``Extragradient method
  in optimization: convergence and complexity,'' \emph{Journal of Optimization
  Theory and Applications}, vol. 176, no.~1, pp. 137--162, 2018.

\bibitem{calafiore2006scenario}
G.~C. Calafiore and M.~C. Campi, ``The scenario approach to robust control
  design,'' \emph{IEEE Transactions on Automatic Control}, vol.~51, no.~5, pp.
  742--753, 2006.

\bibitem{campi2018introduction}
M.~C. Campi and S.~Garatti, \emph{Introduction to the scenario approach}.\hskip
  1em plus 0.5em minus 0.4em\relax SIAM, 2018, vol.~26.

\bibitem{campi2018general}
M.~C. Campi, S.~Garatti, and F.~A. Ramponi, ``A general scenario theory for
  nonconvex optimization and decision making,'' \emph{IEEE Transactions on
  Automatic Control}, vol.~63, no.~12, pp. 4067--4078, 2018.

\bibitem{fabiani2020scenario}
F.~Fabiani, K.~Margellos, and P.~J. Goulart, ``On the robustness of equilibria
  in generalized aggregative games,'' in \emph{2020 59th IEEE Conference on
  Decision and Control (CDC)}.\hskip 1em plus 0.5em minus 0.4em\relax IEEE,
  2020, pp. 3725--3730.

\bibitem{campi2018wait}
M.~C. Campi and S.~Garatti, ``Wait-and-judge scenario optimization,''
  \emph{Mathematical Programming}, vol. 167, no.~1, pp. 155--189, 2018.

\bibitem{nagurney1984comparative}
A.~B. Nagurney, ``Comparative tests of multimodal traffic equilibrium
  methods,'' \emph{Transportation Research Part B: Methodological}, vol.~18,
  no.~6, pp. 469--485, 1984.

\bibitem{brondsted2012introduction}
A.~Brondsted, \emph{An introduction to convex polytopes}.\hskip 1em plus 0.5em
  minus 0.4em\relax Springer Science \& Business Media, 2012, vol.~90.

\bibitem{rockafellar1970convex}
R.~T. Rockafellar, \emph{Convex analysis}.\hskip 1em plus 0.5em minus
  0.4em\relax Princeton University Press, 1970, no.~28.

\end{thebibliography}

\end{document}